\documentclass[twocolumn]{emulateapj} 
\usepackage{amssymb,graphicx} 
\usepackage{epsfig} 
\usepackage{psfrag} 
\usepackage[usenames]{color} 
\usepackage{multirow} 
\usepackage{rotating}

\newcommand{\eqref}[1]{(\ref{#1})} 
 
\shorttitle{NSNS mergers are jet engines}
\shortauthors{Ruiz, Lang, Paschalidis \& Shapiro}

\begin{document} 
 
\title{ 
  Binary neutron star mergers: a jet engine 
  for short gamma-ray bursts} 
 
\author{Milton Ruiz$^{1,2}$, Ryan N.\ Lang$^{1,3}$ ,Vasileios 
  Paschalidis$^4$, Stuart L.\ Shapiro${}^{1,5}$} 
 
\affil{ 
  $^1$Department of Physics, University of Illinois at Urbana-Champaign, 
  Urbana, IL~61801 \\ 
  $^2$Escuela de F{\'\i}ısica, Universidad Industrial de Santander, 
  Ciudad Universitaria, Bucaramanga 680002, Colombia\\ 
  $^3$Leonard E.\ Parker Center for Gravitation, Cosmology, and 
  Astrophysics, University of Wisconsin--Milwaukee, Milwaukee, WI~53211 \\ 
  $^4$Department of Physics, Princeton University, Princeton, NJ~08544\\ 
  $^5$Department of Astronomy \& 
  NCSA, University of Illinois at Urbana-Champaign, Urbana, IL~61801}

 
\begin{abstract} 
  We perform magnetohydrodynamic simulations in full general 
  relativity (GRMHD) of quasi-circular, equal-mass, binary neutron 
  stars that undergo merger. The initial stars are irrotational, $n=1$ 
  polytropes and are magnetized. We explore two types of 
  magnetic-field geometries: one where each star is endowed with a 
  dipole magnetic field extending from the interior into the 
  exterior, as in a pulsar, and the other where the dipole field is 
  initially confined to the interior.  In both cases the adopted 
  magnetic fields are initially dynamically unimportant. The merger 
  outcome is a hypermassive neutron star that undergoes delayed 
  collapse to a black hole (spin parameter $a/M_{\rm BH} \sim 0.74$) 
  immersed in a magnetized accretion disk. About $4000M \sim 
  60(M_{\rm NS}/1.625M_\odot)$ ms following merger, the region above 
  the black hole poles becomes strongly magnetized, and a collimated, 
  mildly relativistic outflow --- an incipient jet --- is 
  launched. The lifetime of the accretion disk, which likely equals 
  the lifetime of the jet, is $\Delta t \sim 0.1 (M_{\rm 
    NS}/1.625M_\odot)$ s. In contrast to black hole--neutron star 
  mergers, we find that incipient jets are launched even when the 
  initial magnetic field is confined to the interior of the 
  stars. 
\end{abstract} 
 
\keywords{black hole physics---gamma-ray burst: general--- 
  gravitation---gravitational waves---stars: neutron} 
\maketitle

 
\section{Introduction} 
 
\begin{figure*} 
\centering 
\includegraphics[clip,trim= 2.5cm 0 3.5cm 0,width=0.18\textwidth,angle=-90,]{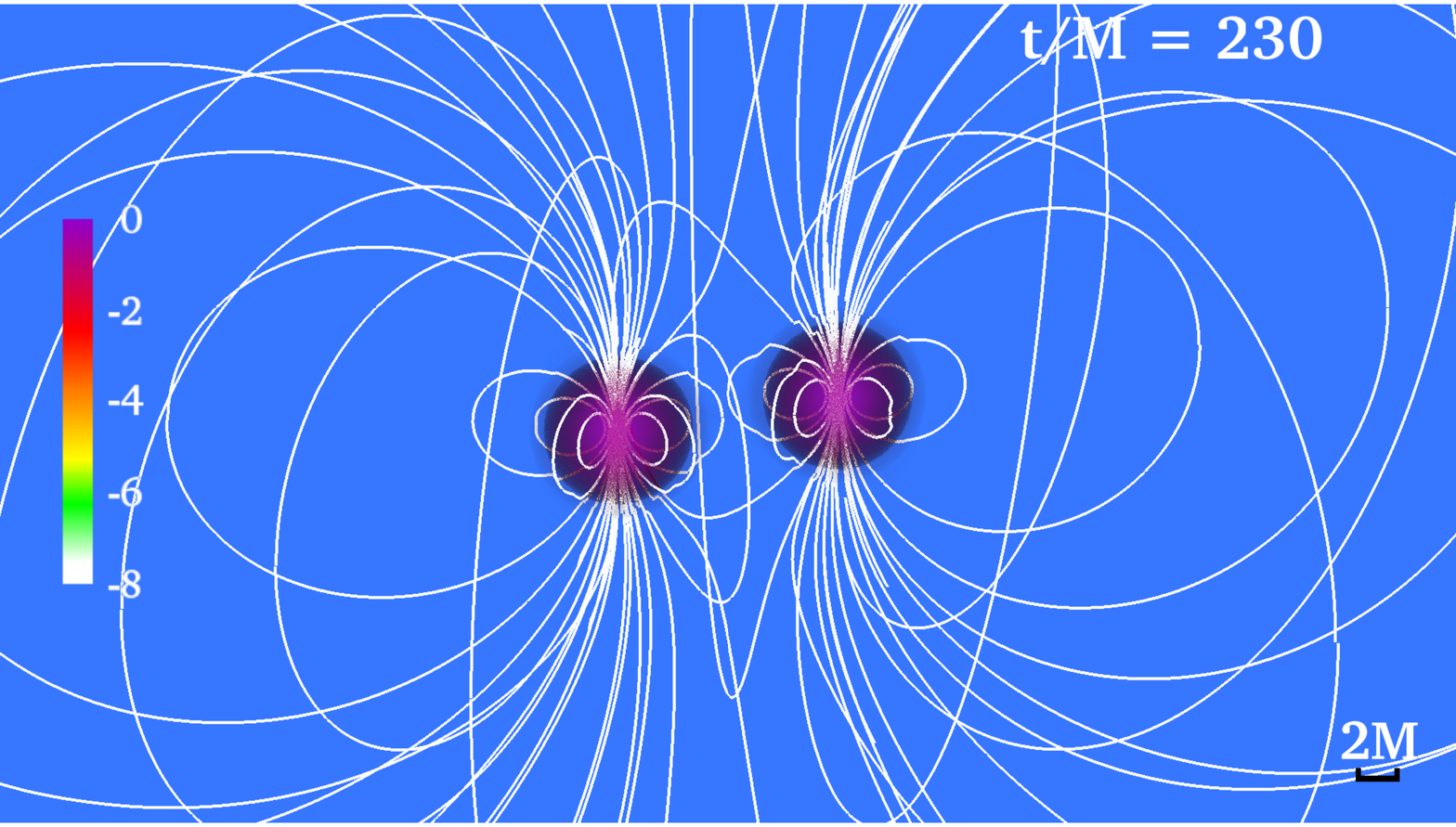} 
\includegraphics[clip,trim= 2.5cm 0 3.5cm 0,width=0.18\textwidth,angle=-90]{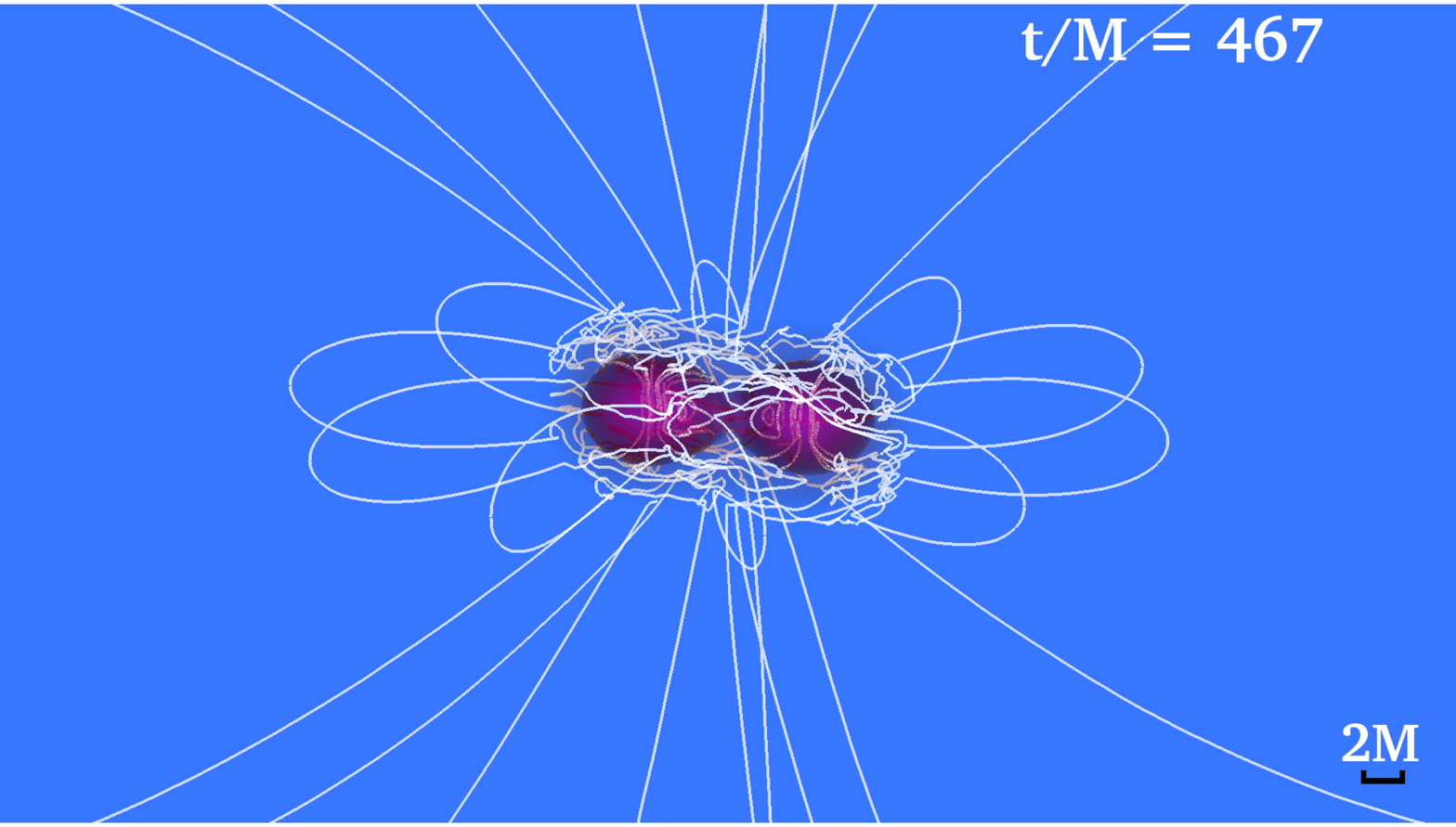} 
\includegraphics[clip,trim= 2.5cm 0 3.5cm 0,width=0.18\textwidth,angle=-90]{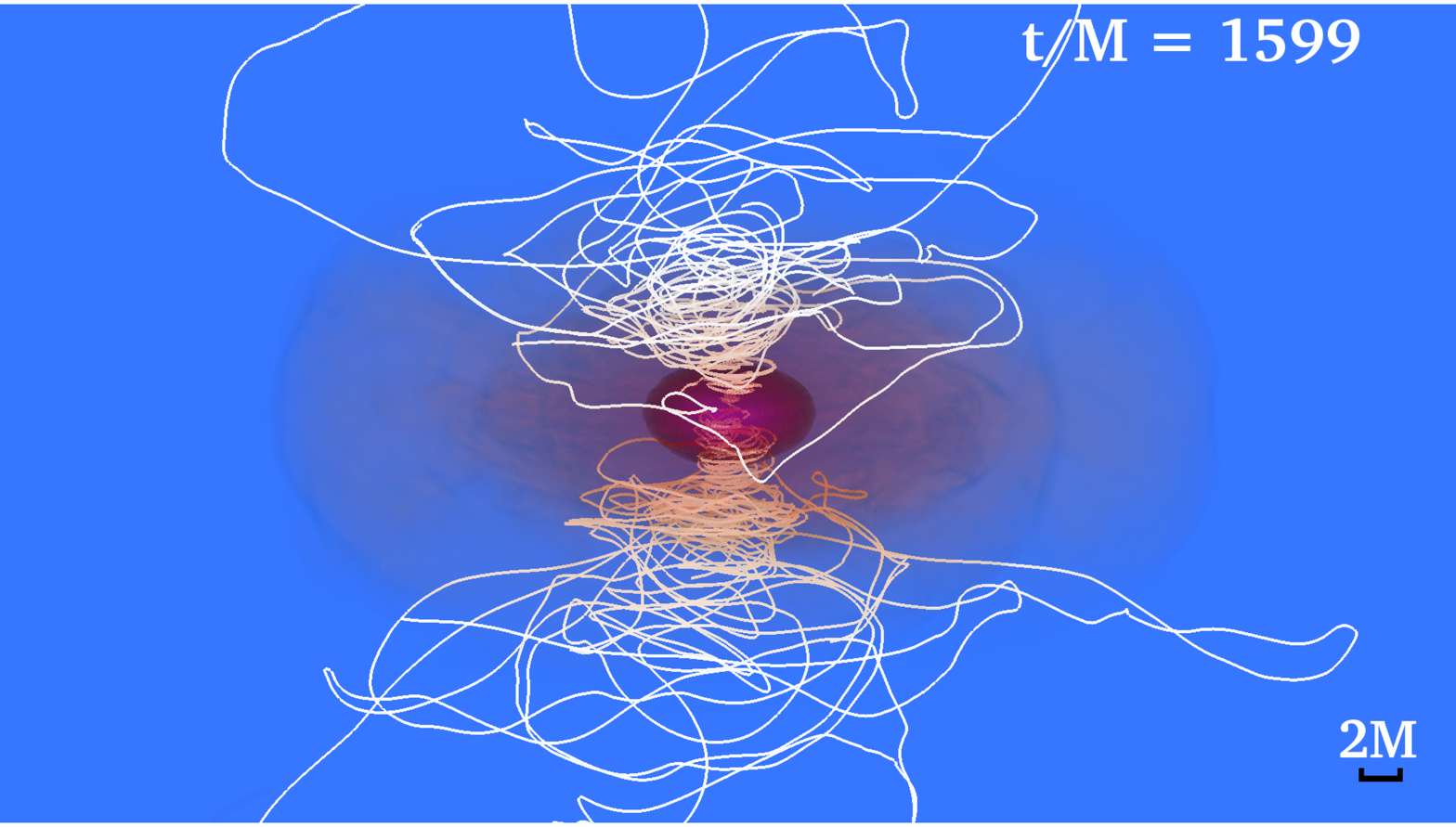} 
\includegraphics[clip,trim= 2.5cm 0 3.5cm 0,width=0.18\textwidth,angle=-90]{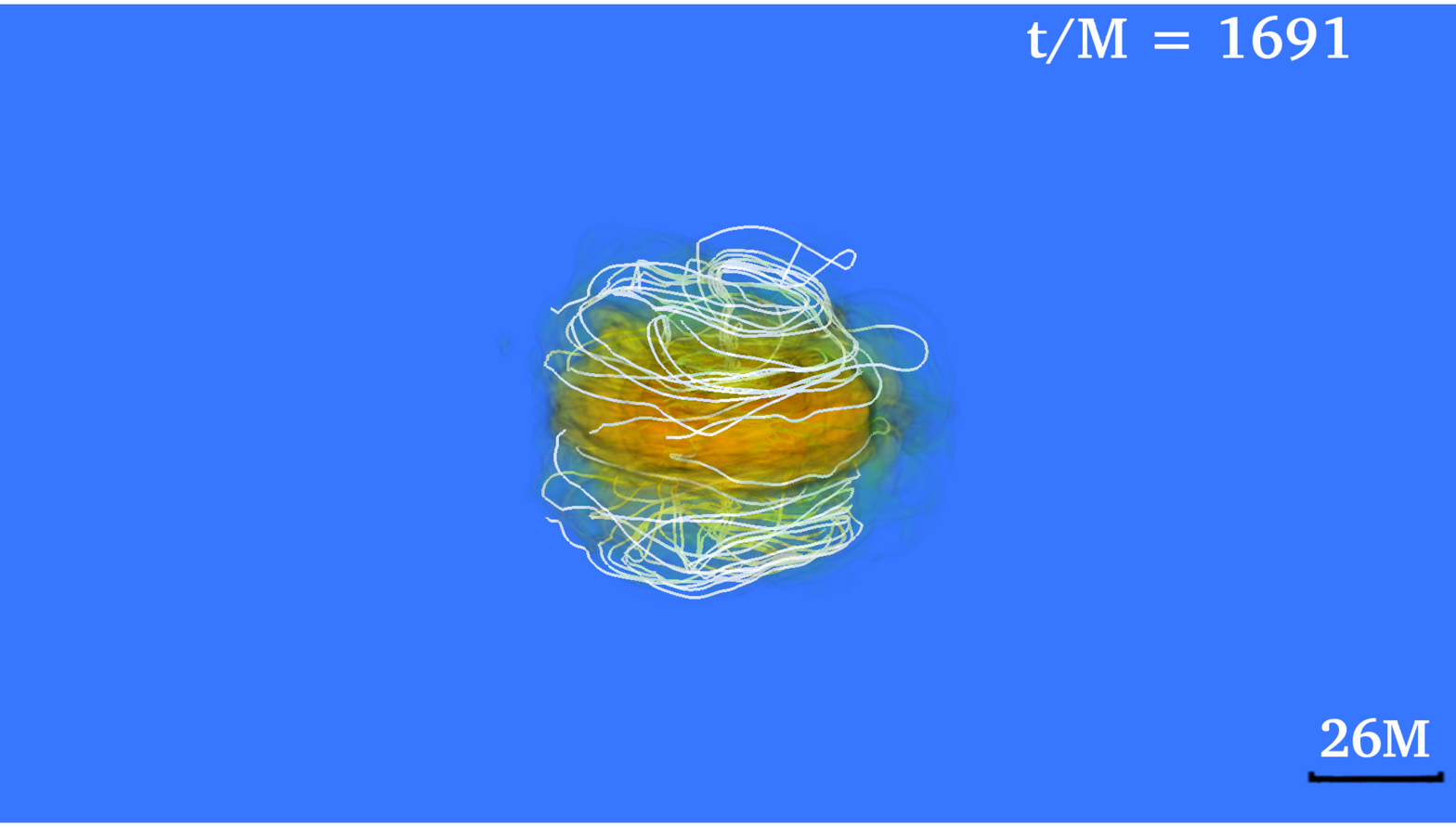} 
\includegraphics[clip,trim= 2.5cm 0 3.5cm 0,width=0.18\textwidth,angle=-90]{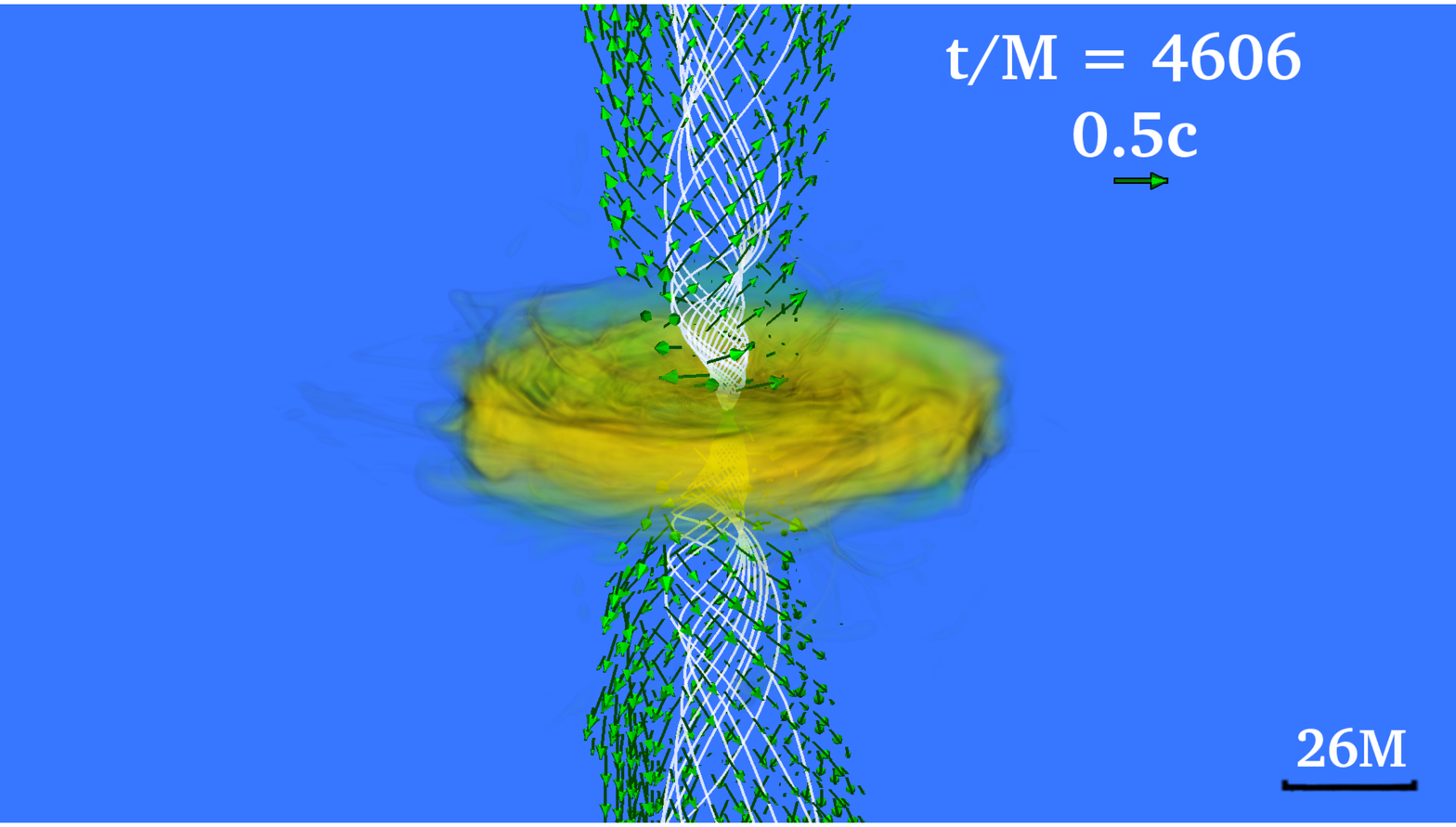} 
\includegraphics[clip,trim= 2.5cm 0 3.5cm 0,width=0.18\textwidth,angle=-90]{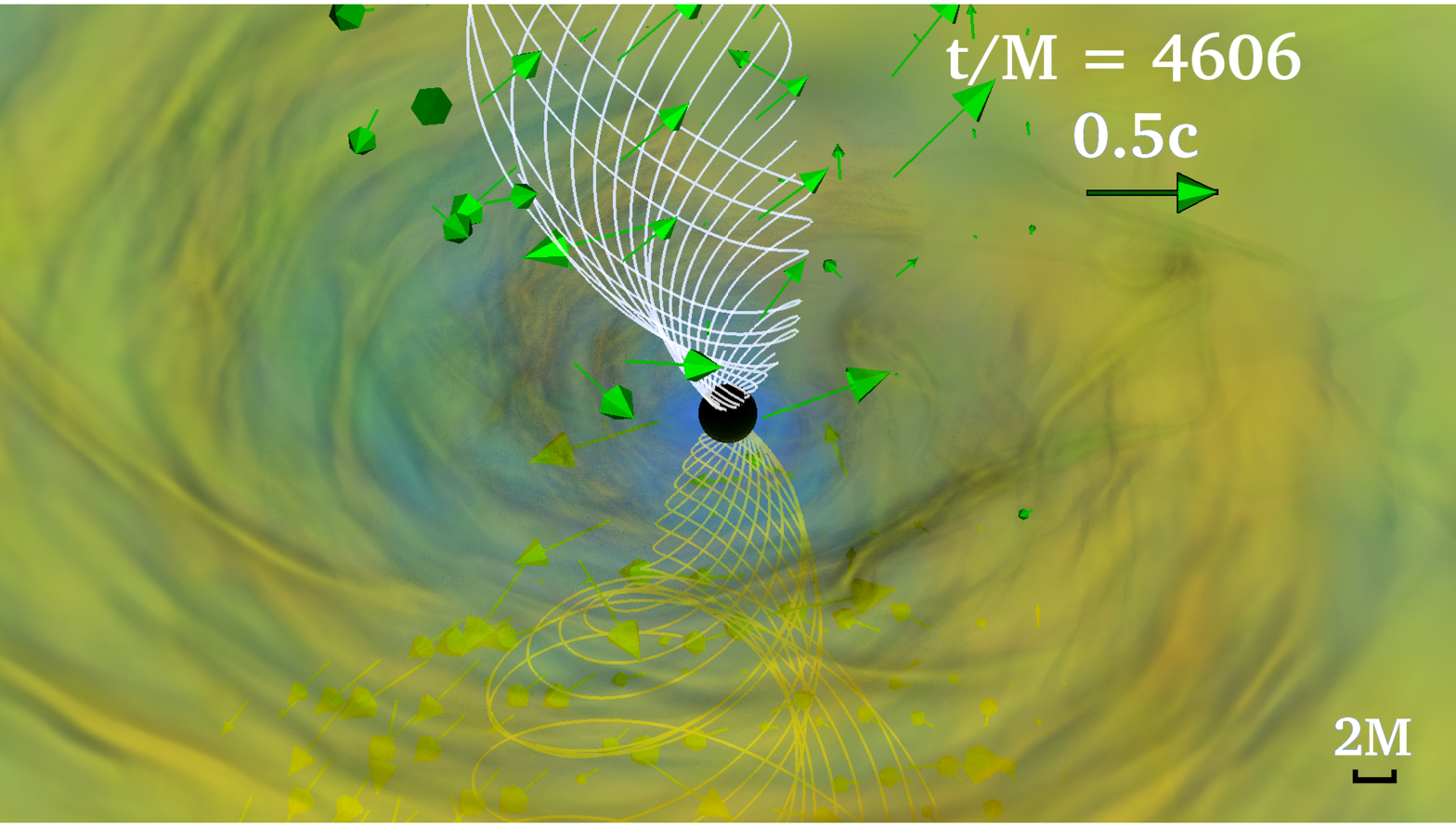} 
\caption{ Snapshots of the rest-mass density, normalized to its 
  initial maximum value $\rho_{0,\text{max}}= 5.9\times 
  10^{14}(1.625\,M_\odot/M_{\rm NS})^2\ \text{g cm}^{-3}$ (log scale), at 
  selected times for the P case.  The arrows indicate plasma velocities, 
  and the white lines show the B-field structure. The bottom 
  middle and right panels highlight 
  the system after an incipient jet is launched. Here $M=1.47\times 
  10^{-2}(M_{\rm NS}/1.625M_\odot)$ ms = $4.43(M_{\rm 
    NS}/1.625M_\odot)$ km. 
\label{fig:snapshots}} 
\end{figure*} 
 
The LIGO and Virgo Collaborations recently reported the first 
direct detection of a gravitational-wave (GW) signal 
and demonstrated that it was produced by the inspiral and 
coalescence  of a binary black hole (BHBH) system
\citep{LIGO_first_direct_GW}. This breakthrough marks the 
beginning of the era of GW astrophysics.  GW signals are 
expected to be generated not only by BHBH binaries, but also by neutron 
star--neutron star (NSNS) and black hole--neutron star (BHNS) binaries. 
 
Merging NSNSs and BHNSs are not only important sources of GWs, but 
also the two most popular candidate progenitors of {\em short} gamma-ray 
bursts (sGRBs)~\citep{EiLiPiSc,NaPaPi,MoHeIsMa,Berger2014}. NSNSs and BHNSs may also 
generate other detectable, transient electromagnetic (EM) signals 
prior to~\citep{Hansen:2000am,ml11,Paschalidis:2013jsa, 
  PalenzuelaLehner2013,2014PhRvD..90d4007P} and 
following~\citep{MetzgerBerger2012,Berger2014,2015MNRAS.446.1115M} merger. 
Combining GW and EM signals from these mergers could test relativistic 
gravity and constrain the NS equation of state (EOS). Moreover, 
an association of a GW event with an sGRB (the holy grail of 
``multimessenger astronomy'') would provide convincing evidence for 
the compact binary coalescence model.  However, the interpretation of 
EM and GW signals from such mergers will rely on a theoretical 
understanding of these events, which requires simulations in full 
general relativity  to treat the strong dynamical fields and high 
velocities arising in these scenarios. There have been multiple 
studies of compact binary mergers.  For NSNSs, see~\citet{faber_review} 
for a review and~\cite{Paschalidis:2012ff,gold,East2012NSNS,
Neilsen:2014hha,Dionysopoulou2015,Sekiguchi2015,Dietrich2015}; 
and \cite{Palenzuela2015} for recent results. These earlier studies 
have advanced our knowledge of EOS effects, neutrino transport, ejecta 
properties, and magnetospheric phenomena. However, few studies focused 
on the potential for NSNSs to power sGRBs. 
 
Recent work by~\cite{prs15} (hereafter PRS) demonstrated that mergers 
of magnetized BHNS systems can launch jets and be the engines that 
power sGRBs. The key ingredient for generating a jet was found to be 
the initial endowment of the NS with a dipole B-field that extends 
into the NS exterior as in a pulsar magnetosphere. By contrast, if 
the initial magnetic field is confined to the interior of the NS, no 
jet is observed~\citep{Etienne:2012te,kskstw15}. 
 
The primary motivation for this paper is to answer the question: 
{\it Can NSNS mergers produce jets in the same way as BHNS systems, or 
  does this mechanism require an initial BH?}  Recently, it was shown 
that neutrino annihilation may not be strong enough to power 
jets~\citep{jojb15}, so MHD processes must play a major role for jet 
formation. Previous ideal GRMHD simulations by~\cite{rgbgka11} suggest 
that NSNS mergers may launch a relativistic jet, while those 
by~\cite{kkssw14}, which focus on different initial configurations, show 
otherwise. Both of these studies have considered only scenarios where 
the B-field is initially confined to the {\em interior} of the two 
NSs. 
 
Here, we describe the results of ideal GRMHD simulations of NSNSs in 
which we follow PRS and allow an initially strong, but dynamically 
unimportant dipole B-field to extend from the interior of the NSs 
into the exterior.  We call this configuration the pulsar model 
(hereafter P). The existence of pulsars suggests that this may be the 
astrophysically most common case. As in PRS, we define an {\em incipient jet} 
as a collimated, mildly relativistic outflow which is at least partially 
magnetically dominated ($b^2/(2\rho_0)>1$, where $b^2 = B^2/4 \pi$ and 
$\rho_0$ is the rest-mass density).  We find that the P configuration 
launches an incipient jet. To study the impact of the initial B-field 
geometry and to compare it with previous studies, we also perform simulations 
where the field is confined to the interior of the NSs (hereafter the I 
model), keeping its strength at the stellar center the same as in the P case. 
In contrast to BHNS systems, we find that interior-only initial B-fields 
also lead to jet formation in NSNSs.  Throughout this work, 
geometrized units ($G=c=1$) are adopted unless otherwise 
specified. 
 
 
\section{Methods} 
We use the Illinois GRMHD code, which is built on the {\tt 
  Cactus\footnote{http://www.cactuscode.org}} infrastructure and uses 
the {\tt Carpet\footnote{http://www.carpetcode.org}} code for adaptive 
mesh refinement.  We use the {\tt AHFinderDirect} 
thorn~\citep{ahfinderdirect} to locate apparent horizons. 
This code has been thoroughly tested and used in the past in different 
scenarios involving magnetized compact binaries~(see, 
e.g.,~\cite{Etienne:2007jg,lset08,Etienne:2012te,Gold:2013zma}; and 
\cite{Gold2014}). 
For implementation details, see~\cite{Etienne:2010ui,Etienne:2011ea} and 
\cite{Farris:2012ux}. 
 
In all simulations we use seven levels of refinement with two sets of 
nested refinement boxes (one for each NS) differing in size and 
resolution by factors of two. The finest box around each NS has a 
half-side length of $\sim 1.3\,R_{\rm NS}$, where $R_{\rm NS}$ is the 
initial NS radius.  For the I model, we run simulations at two 
different resolutions: a ``normal'' resolution (model IN), in which 
the finest refinement level has grid spacing $0.05 M = 227(M_{\rm 
  NS}/1.625M_\odot)$ m, and a ``high'' resolution (model IH), 
in which the finest level has spacing 
$0.03 M = 152(M_{\rm NS}/1.625M_\odot)$ m. For the P model, we always use 
the high resolution.  These choices resolve the 
initial NS equatorial diameter by $\sim 120$ and $\sim 180$ points, 
respectively. In terms of grid points per NS diameter, our high 
resolution is close to the medium resolution used in \cite{kkssw14}, 
which covered the initial stellar diameters by $\sim 205$ points. We 
set the outer boundary at $245M\approx 1088(M_{\rm 
  NS}/1.625M_{\odot})$ km and impose reflection symmetry across the 
orbital plane. 
 
\setlength{\tabcolsep}{2pt} 
\begin{deluxetable}{ccccccc} 
\scriptsize 
\tablewidth{0pt} 
\tablecaption{Summary of results \label{tab:models_NSNS}} 
\tablecolumns{6} 
\tablehead{ 
  \colhead{Case} 
& \colhead{$\Gamma_L$ \tablenotemark{a}} 
& \colhead{$B_{\text{rms}}$ \tablenotemark{b}} 
& \colhead{$\dot{M}$ \tablenotemark{c}} 
& \colhead{$M_{\rm disk}/M_{\rm 0}$ \tablenotemark{d}} 
& \colhead{$\tau_{\rm disk}$ \tablenotemark{e}} 
& \colhead{$L_{\rm EM}$ \tablenotemark{f}} 
} 
\startdata 
P$\ $      &    1.25 &$10^{16.0}$  &  0.33  & 1.0\%   & 0.10   & $10^{51.3}$  \\ 
IN         &    1.21 &$10^{15.8}$  &  0.54  & 1.1\%   & 0.07   & $10^{50.9}$  \\ 
IH         &    1.15 &$10^{15.7}$  &  0.77  & 1.5\%   & 0.06   & $10^{50.7}$  \\
\enddata 
\tablenotetext{a}{Maximum fluid Lorentz factor near the end of the simulation.} 
\tablenotetext{b}{rms value of the HMNS B-field before collapse.} 
\tablenotetext{c}{Rest-mass accretion rate in units of $M_\odot\rm{s}^{-1}$ 
at $t-t_{\rm BH} = 2100M \approx 31(M_{\rm NS}/1.625M_\odot)$ ms.} 
\tablenotetext{d}{Ratio of disk rest mass to initial total rest mass 
at $t-t_{\rm BH} = 2100M \approx 31(M_{\rm NS}/1.625M_\odot)$ ms.} 
\tablenotetext{e}{Disk lifetime $M_{\rm disk}/\dot M$ in $(M_{\rm NS}/1.625M_\odot)$ s.} 
\tablenotetext{f}{Poynting luminosity in erg s$^{-1}$, 
  time-averaged over the last $500M \sim 7.4(M_{\rm NS}/1.625M_\odot)$ 
ms of the evolution after the jet is well--developed.} 
\end{deluxetable} 
 
The quasi-equilibrium NSNS initial data were generated with 
the {\tt LORENE} libraries.\footnote{http://www.lorene.obspm.fr} 
Specifically, we use the $n=1$, irrotational case listed in 
\cite{tg02}, Table III, $M/R = 0.14$ vs. $0.14$, row $3$, for 
which the rest mass of each NS is $1.625M_{\odot} (k/269.6
\mathrm{km^2})^{1/2}$, with $k$ the polytropic constant. This same case was 
used in~\cite{rgbgka11}. As in PRS, we evolve the initial data up to the final 
two orbits prior to merger ($t=t_B$), at which point each NS is seeded 
with a dynamically unimportant B-field following one of two prescriptions: 

\begin{itemize}
\item[(1)] The P case (Figure~\ref{fig:snapshots}, upper left), for which we use a 
dipole B-field corresponding to Eq. (2) in~\cite{Paschalidis:2013jsa}.  We 
choose the parameters $I_0$ and $r_0$ such that the magnetic-to-gas-pressure 
ratio at the stellar center is $\beta^{-1}=0.003125$. The resulting B-field 
strength at the NS pole measured by a normal observer is 
${B}_{\rm pole}\simeq 1.75\times 10^{15}(1.625M_\odot/M_{\rm NS})$ G. While 
this B-field is astrophysically 
large, we choose it so that following merger, the rms value of the 
field strength in the hypermassive neutron star (HMNS) remnant, is 
close to the values found in recent very-high-resolution 
simulations~\citep{Kiuchi:2015sga}, which showed that the 
Kelvin-Helmholtz instability (KHI) during merger can boost the rms 
B-field to $10^{15.5}$G with local values reaching even $10^{17}$G.  Our choice of the 
B-field strength thus provides an ``existence proof'' for jet 
launching following NSNS mergers with the finite computational 
resources at our disposal. To capture the evolution of the exterior B-field
in this case and simultaneously mimic force-free conditions that 
likely characterize the exterior, we follow PRS and set a variable-density 
atmosphere at $t=t_B$ such that the exterior plasma parameter 
$\beta_{\rm ext}=0.01$. This variable-density prescription, imposed at 
$t=t_B$ \emph{only}, is expected to have no impact on the 
outcome~(cf. PRS). With our choice of~$\beta_{\rm ext}$, the 
amount of total rest mass does not increase by more than~$\sim 0.5\%$. 
 
\item[(2)] The I case, which also uses a dipole field but 
confines it to the interior. We generate the vector potential through 
Eqs.\ (11), (12) in~\cite{Etienne:2011ea}, choosing $P_\mathrm{cut}$ to be 
$1\%$ of the maximum pressure, $n_b =2$, and $A_b$ such that 
the strength of the B-field at the stellar center coincides with that 
in the P case. Unlike the P case, a variable-density atmosphere is 
not necessary, so we use a standard constant-density 
atmosphere with rest-mass density $10^{-10}\rho_{0,\rm max}$, where 
$\rho_{0,\rm max}$ is the initial maximum value of the rest-mass 
density. 
\end{itemize}

In both the P and I cases, the magnetic dipole moments are aligned with 
the orbital angular momentum. During the evolution, we impose a 
floor rest-mass density of $10^{-10}\rho_{0,\rm max}$.  We employ a 
$\Gamma$-law EOS, $P=\rho_0\epsilon$, with $\epsilon$ the specific 
internal energy, and allow for shocks. 
 
 
\section{Results}

\begin{figure} 
  \centering 
    \includegraphics[width=0.40\textwidth,trim=0 0 0 0,clip=true,angle=-90]{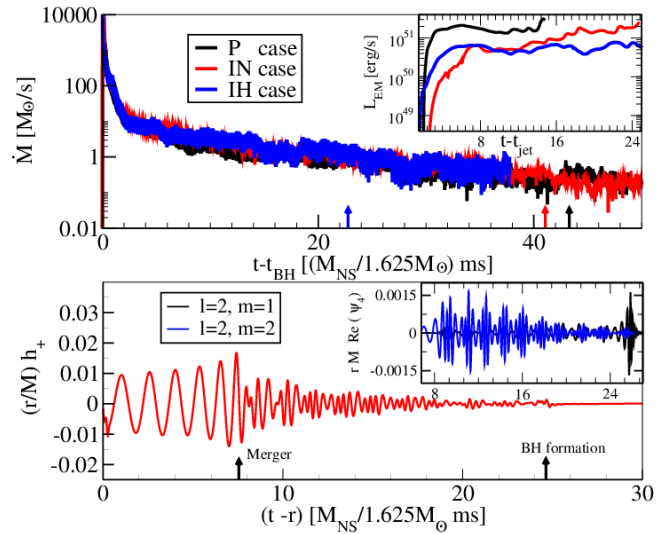} 
    \caption{Top: Rest-mass accretion rate $\dot{M}$. The arrows 
      indicate the time ($t_{\rm jet}$) at which the incipient jet reaches 
      $z=100M$. The inset shows the outgoing EM luminosity for $t>t_{\rm jet}$ 
      computed on a coordinate sphere of radius $r=115M\approx 510(M_{\rm NS}/1.625M_\odot)$ km. 
      Bottom: GW amplitude $h_+$ vs. retarded time 
      for the P case. The inset focuses on the post-merger phase, 
      showing the $(l,m)=(2,2)$ and $(l,m)=(2,1)$ (multiplied by 20) modes 
      of the Newman--Penrose scalar $\Psi_4$. 
  \label{fig:accretion_waveform}} 
\end{figure}

The outcomes and basic dynamics for all our cases are similar, hence we 
show snapshots and discuss the evolution only for the P 
case, unless otherwise specified. All coordinate times will 
refer to the P case too, unless otherwise specified. We summarize 
key results for all cases in Table~\ref{tab:models_NSNS}. 
 
All cases evolve the same until merger. The two NSs orbit each other 
with their B-fields frozen in.  Gravitational-radiation loss causes 
the orbit to shrink, and the NSs make contact at $t=t_{\rm 
  merger}\approx 465M\sim 3.5(M_{\rm NS}/1.625M_\odot)$ ms, when the 
stars are oblate due to tidal effects  (Fig.~\ref{fig:snapshots}, 
upper middle). Then a double-core remnant forms with the two dense 
cores rotating about each other and gradually coalescing  
(Fig.~\ref{fig:snapshots}, upper right).  

Just before the HMNS collapses, the rms 
B-field strength is $\sim 10^{15.7}-10^{16}$ G (see 
Table~\ref{tab:models_NSNS}), a little larger than the 
values~\cite{Kiuchi:2015sga} reported, but consistent with 
\cite{2013ApJ...769L..29Z}. 
 
During the HMNS stage from $t-t_{\rm merger}\approx 115M\sim 1.7(M_{\rm 
  NS}/1.625M_\odot)$ ms up until BH formation, there is no 
significant enhancement of the B-field. This result is anticipated 
because we chose our initial B-fields such that they are near 
saturation ($\beta \sim 100-1000$) following merger. We find that 
inside the HMNS we resolve the wavelength $\lambda_{\rm MRI}$ of the 
fastest-growing magneto-rotational-instability (MRI) mode by more than 
$10$ points.  In addition, $\lambda_{\rm MRI}$ fits within the HMNS, 
so we conclude that MRI-induced turbulence is operating during this stage. 
As the total rest mass of the HMNS exceeds the maximum value allowed by 
uniform rotation, i.e. $\sim 2.4M_\odot(k/296.6\ \mathrm{km^2})^{1/2}$ for 
$\Gamma=2$~\citep{LBS2003ApJ}, it undergoes delayed collapse to a 
BH~\citep{BaShSh,dlsss06a} at $t-t_{\rm merger} \approx 1215M\sim 
18(M_{\rm NS}/1.625M_\odot)$ ms, in both the P and IN cases. In the 
IH case, collapse takes place later at $t-t_{\rm merger}\approx 
2135M\sim 31.5(M_{\rm NS}/1.625M_\odot)$ ms. The sensitivity of the 
collapse time for short-lived HMNSs to the B-field is physical and has 
been observed previously~\citep{grb11}.  Its dependence on resolution, 
even in purely hydrodynamic simulations, has been noted as 
well~\citep{PEFS2015,PEFS2016}.  In all cases the BH has a mass of 
$M_{\rm BH}\approx 2.85M_\odot(M_{\rm NS}/1.625M_\odot)$ with spin 
$a/M\simeq 0.74$, at high resolution, or $a/M\simeq 0.8$, at normal 
resolution.

Shortly after BH formation, plasma velocities above the BH poles point 
toward the BH (apart from some material 
ejected during merger), but the winding of the B-field above the BH 
poles is well underway (Fig.~\ref{fig:snapshots}, bottom left). By 
$t-t_{\rm BH} \approx 2800M \sim 41 (M_{\rm NS}/1.625M_\odot)$ ms, 
where $t_{\rm BH}$ is the BH formation time, the B field has been 
wound into a helical funnel (Fig.~\ref{fig:snapshots}, bottom middle and right). 
Unlike in the BHNS case of PRS, the B-field does not grow following BH
formation: the existence of the HMNS 
phase instead allows the B-field to build to saturation levels prior 
to BH formation. We do observe a gradual growth in $b^2/(2\rho_0)$ 
above the BH pole from $\sim 1$ to $\sim 100$, due to the emptying of 
the funnel as matter accretes onto the BH. At 
$t-t_{\rm BH}\approx 2000M \sim 30 (M_{\rm NS}/1.625M_\odot)$ ms, the 
fluid velocities begin to turn around and point outward. At 
$t-t_{\rm BH}\approx 2900 M\sim 43 (M_{\rm NS}/1.625M_\odot)$ ms, the 
outflow extends to heights greater than $100M \sim 445 (M_{\rm NS}/1.625 
M_\odot)$ km. At this point an incipient jet has formed 
(Fig.~\ref{fig:snapshots}, bottom middle and right).  We observe similar 
evolution in the I cases, although in the IH case the incipient jet 
is well--developed by $t-t_{\rm BH}\sim 1550M\sim 23 (M_{\rm NS}/1.625 
M_\odot)$ ms, while in the IN case it takes $t-t_{\rm BH}\sim 
2850M\sim 42 (M_{\rm NS}/1.625M_\odot)$ ms for the incipient jet 
to be launched (see the arrows in Fig.\ \ref{fig:accretion_waveform}). 
 
 
\begin{figure} 
\centering 
\includegraphics[width=0.35\textwidth,trim=0 0 0 0,clip=true,angle=-90]{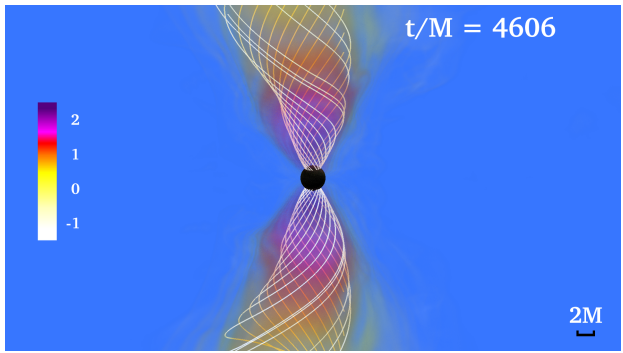} 
\caption{ 
\label{fig:b2_2rho3D} 
Ratio~$b^2/(2\rho_0)$ (log scale) at $t-t_{\rm merger}\approx 
4135M\sim 61(M_{\rm NS}/1.625M_\odot)$ ms for the P case. The white 
lines indicate the B-field lines plotted in the funnel 
where~$b^2/(2\rho_0) \geq 10^{-2}$.  Magnetically dominated 
areas~($b^2/(2\rho_0)\geq 1$) extend to heights greater than $20M\approx 
20\,r_{\rm BH}$ above the BH horizon (the black sphere). Here $r_{\rm 
  BH}=2.2(M_{\rm NS}/1.625M_\odot)$ km.} 
\end{figure}
 
In all cases, the accretion rate $\dot{M}$ begins to settle into a 
steady state at $t-t_{\rm BH}\approx 1350 M\sim 20(M_{\rm NS}/1.625 
M_\odot)$ ms and slowly decays thereafter (Fig.\ \ref{fig:accretion_waveform}, 
top). At $t-t_{\rm BH}\approx 2100 M\sim 
31(M_{\rm NS}/1.625M_\odot)$ ms, the accretion rate is roughly 
$0.33\,M_\odot\mathrm{s}^{-1}$, at which time the disk mass is $~\sim 0.05\,M_\odot 
(M_{\rm NS}/1.625M_\odot)$. Thus, the disk will be accreted in $\Delta 
t\sim M_{\rm disk}/\dot{M}\sim 0.1$ s, implying that the 
engine's fuel will be exhausted on a time scale consistent with the 
duration of the very short-duration sGRBs (see, 
e.g.,~\citet{Kann:2008zg}). 
 
In the BH--accretion disk system, we again resolve $\lambda_{MRI}$ by 
more than $10$ grid points. For the most part, $\lambda_{MRI}$ fits 
inside the disk. MRI is thus operating in these simulations, but it is 
saturated, with the toroidal and poloidal B-field components 
approximately equal in magnitude and not growing. Near the end of the 
simulations the B-field magnitude above the BH pole is 
$\sim 10^{16}$ G. 
 
As in PRS, we have tracked the motion of individual Lagrangian fluid 
tracers. We find that they follow the expected helical motion 
of the outflow and that the gas pouring into the funnel and comprising 
the outflow originates from the disk.  This test confirms the physical nature 
of the jet. 
 
During the inspiral and throughout BH formation, GWs are emitted with the 
$(l,m)=(2,2)$ mode being dominant. This includes quasi-periodic 
spindown GWs from the transient HMNS (\cite{ShibataPhysRevLett.94.201101}; 
for a review and references, see \cite{baumgartebook10}). 
However, we find that $(l,m)=(2,1)$ 
GW modes also develop during and after merger (Fig.\ \ref{fig:accretion_waveform}, 
bottom). While we see no evidence for a one-arm 
instability~\citep{PEFS2015}, $m=1$ modes 
may be generated from asymmetries in the two cores following 
merger. Like post-merger $m=2$ modes, $m=1$ modes due to 
asymmetries in the two cores of double-core HMNSs may be used 
to constrain the NS EOS.

Fig.~\ref{fig:b2_2rho3D} displays the ratio $b^2/(2\rho_0)$ at 
$t-t_{\rm merger}\approx 4200{\rm M}\sim 68(M_{\rm NS}/1.625 
M_\odot)$ ms. Magnetically dominated areas ($b^2/(2\rho_0)>1$) extend 
to heights $\gtrsim 20M\approx 20\,r_{\rm BH}$ above the BH, where 
$r_{\rm BH}$ is the BH apparent horizon radius. Using the 
$b^2/(2\rho_0) \sim 10^{-2}$ contour as a rough definition for the 
funnel boundary, the funnel opening half-angle is $\sim 20^\circ-30^\circ$ 
degrees, which is consistent with PRS. The maximum value of the Lorentz factor 
inside the funnel is $\Gamma_L \sim 1.1-1.25$ 
(see~Table~\ref{tab:models_NSNS}), implying only a mildly relativistic 
flow. However, steady-state and axisymmetric jet 
models~\citep{B2_over_2RHO_yields_target_Lorentz_factor} show that the 
maximum attainable value of $\Gamma_L$ is approximately equal to 
$b^2/(2\rho_0$), which reaches values $\gtrsim 100$ within the 
funnel. Hence, these incipient jets may be accelerated to $\Gamma_L 
\sim 100$, as is necessary to explain sGRB phenomenology.  (However, 
our code may not be reliable at values of $b^2/(2\rho_0)>100$.) In 
the asymptotically flat region at heights greater than 
$100M \sim 650(M_{\rm NS}/1.625\,M_\odot)$ km, the specific energy $E=-u_0-1>0$ 
is always positive and therefore the outflowing plasma is unbound. 
 
To assess if the Blandford-Znajek (BZ) effect~\citep{BZeffect} is 
operating in our system as in PRS, we measure the ratio of the angular 
frequency of the B-field, $\Omega_F$, 
to the angular frequency of the BH, $\Omega_H$. 
On an $x-z$ meridional slice passing through the BH centroid and along 
coordinate semicircles of radii $r_{\rm BH}$ and $2r_{\rm BH}$, we 
find that $\Omega_F/\Omega_H\sim 0.1-0.25$ within an opening angle of 
$20^\circ$ degrees from the BH rotation axis, within which the force-free BZ 
solution might apply. The observed deviation of $\Omega_F/\Omega_H$ 
from the split-monopole value $\sim 0.5$ can be due either to the 
gauge in which we compute $\Omega_F$, deviations from a split-monopole 
B-field, or  possibly from inadequate resolution.  The outgoing Poynting 
luminosity is $L_{EM}\sim 10^{50.7}-10^{51.3}$ erg s$^{-1}$ (see 
Fig.~\ref{fig:accretion_waveform}, Table~\ref{tab:models_NSNS}), which is a only a little less than the 
EM power expected from the BZ effect~\citep{Thorne86}: $L_{EM} \sim 
10^{52}[(a/M_{\rm BH})/0.75]^2(M_{\rm BH}/2.8 M_\odot)^2(B/10^{16} 
{\rm G})^2$erg s$^{-1}$.  The BZ effect is likely operating here, but 
higher resolution may be necessary to match the expected luminosity 
more closely. 
 
\section{Conclusions} 
 
We showed that an NSNS system with an initially high but 
dynamically weak B-field, launches an incipient jet (an unbound, 
collimated, and mildly relativistic outflow). This occurs following the delayed 
collapse of the HMNS, both for initially dipole B-fields which extend 
from the NS interior into its exterior and initially dipole B-fields 
confined to the interior. This last result differentiates NSNSs from 
BHNSs. The accretion timescale of the remnant disk and energy output 
are consistent with very short sGRBs, {demonstrating} that 
NSNSs can indeed provide the central engines that power such phenomena. 
Our results were obtained with high initial B-fields, 
which match the post-merger expectations from B-field amplification 
due to the KHI. However, to capture the KHI-induced B-field growth, 
resolutions about five times higher than those adopted here are 
necessary. {Evolving} for $4000M \sim 60(M_{\rm NS}/1.625 
M_\odot)$ ms at such high resolution would take years with current 
resources. The simulations presented here show that if the B-fields
in NSNS mergers can be amplified by the KHI to $\gtrsim 10^{15.5}$ G, 
then these systems are viable sGRB engines. We anticipate, 
but have not yet proved, that jets can be launched even if one starts 
with B-field strengths of order $10^{12}$ G. We also expect that the 
time interval between merger and jet launch will be longer the weaker 
the initial B-field. 

Different EOSs affect details such as the amount of disk mass, as well
as the ejection of different amounts of matter (see
e.g.~\cite{Hotokezaka13,Palenzuela:2015dqa}), and therefore the ram
pressure of the atmosphere surrounding the remnant BH-disk system.
The above may be the main reason for the non-observance of an
incipient jet in the NSNS simulations reported by~\cite{kkssw14}. We
infer from their description that the merger with their adopted EOS
ejects more matter in the atmosphere with longer fall-back time than
in our cases. For the outflow to emerge, the B-field must overcome the
ram pressure of the infalling matter. The only way for this to happen
is if the ambient matter density decreases with time, which occurs
because the atmosphere becomes thinner as it accretes onto the
BH. \cite{kkssw14} have reported that at $t-t_{BH}\sim 26$ ms, there
still is matter in the atmosphere that is accreting, hence the ram
pressure is larger than the B-field pressure. It is likely that their
calculations require longer integration times for an incipient jet to
emerge.

Recently, NSNS simulations with a weaker interior-only initial B-field
were carried out by~\cite{Endrizzi:2016kkf}. Their resolution was too
low to capture MRI or the KHI, hence the B-field did not amplify
appreciably and as a result, no jets were observed.

A few caveats remain. First, several quantities are sensitive to 
resolution,  such as the HMNS lifetime and the 
exact EM Poynting luminosity. However, other evolution 
characteristics, such as HMNS formation following NSNS merger, delayed 
collapse, the remnant BH mass and spin, the disk mass, and the ultimate 
emergence of the incipient jet are almost insensitive to 
resolution. The above conclusions suggest that higher resolutions than 
those adopted here are necessary for very accurate calculations, but 
the essential nature of incipient jet emergence is robust.  Second, 
there is microphysics that we do not model here, such as the effects 
of a realistic hot, nuclear EOS and neutrino transport. 
We plan to implement such processes and address all of the aforementioned 
issues in future studies.

 
\acknowledgements We thank the Illinois Relativity Group REU team  
members Sean E. Connelly, Cunwei Fan, Abid Khan, and Patchara Wongsutthikoson 
for their assistance in creating Figs.~\ref{fig:snapshots} and~\ref{fig:b2_2rho3D}. 
This work has been supported in part by NSF grant PHY-1300903 and NASA 
grant NNX13AH44G at the University of Illinois at Urbana-Champaign.  M.R. was also
supported in part by Colciencias under program ``Es tiempo de Volver.'' 
R.L. acknowledges support from a Fortner Fellowship at UIUC as well as NSF 
grant PHY-1307429.  V.P. acknowledges support from the Simons foundation and 
NSF grant PHY-1305682.  This work made use of the Extreme Science and Engineering 
Discovery Environment (XSEDE),  which is supported by National Science Foundation 
grant number ACI-1053575. This research is part of the Blue Waters sustained-petascale 
computing project, which is supported by the National Science Foundation 
(awards OCI-0725070 and ACI-1238993) and the state of Illinois. Blue Waters is 
a joint effort of the University of Illinois at 
Urbana-Champaign and its National Center for Supercomputing Applications.

\bibliographystyle{hapj} 
\bibliography{references}

\begin{thebibliography}{53}
\expandafter\ifx\csname natexlab\endcsname\relax\def\natexlab#1{#1}\fi

\bibitem[{Abbott {et~al.}(2016)Abbott, Abbott, Abbott, Abernathy, Acernese,
  Ackley, Adams, Adams, Addesso, Adhikari, Adya, Affeldt, Agathos, Agatsuma,
  Aggarwal, Aguiar, Aiello, Ain, Ajith, Allen, Allocca, Altin, Anderson,
  Anderson, Arai, Arain, Araya, Arceneaux, Areeda, Arnaud, Arun, Ascenzi,
  Ashton, Ast, Aston, Astone, Aufmuth, Aulbert, Babak, Bacon, Bader, Baker,
  Baldaccini, Ballardin, Ballmer, Barayoga, Barclay, Barish, Barker, Barone,
  Barr, Barsotti, Barsuglia, Barta, Bartlett, Barton, Bartos, Bassiri, Basti,
  Batch, Baune, Bavigadda, Bazzan, Behnke, Bejger, Belczynski, Bell, Bell,
  Berger, Bergman, Bergmann, Berry, Bersanetti, Bertolini, Betzwieser, Bhagwat,
  Bhandare, Bilenko, Billingsley, Birch, Birney, Birnholtz, Biscans, Bisht,
  Bitossi, Biwer, Bizouard, Blackburn, Blair, Blair, Blair, Bloemen, Bock,
  Bodiya, Boer, Bogaert, Bogan, Bohe, Bojtos, Bond, Bondu, Bonnand, Boom, Bork,
  Boschi, Bose, Bouffanais, Bozzi, Bradaschia, Brady, Braginsky, Branchesi,
  Brau, Briant, Brillet, Brinkmann, Brisson, Brockill, Brooks, Brown, Brown,
  Brown, Buchanan, Buikema, Bulik, Bulten, Buonanno, Buskulic, Buy, Byer,
  Cabero, Cadonati, Cagnoli, Cahillane, Bustillo, Callister, Calloni, Camp,
  Cannon, Cao, Capano, Capocasa, Carbognani, Caride, Diaz, Casentini, Caudill,
  Cavagli\`a, Cavalier, Cavalieri, Cella, Cepeda, Baiardi, Cerretani, Cesarini,
  Chakraborty, Chalermsongsak, Chamberlin, Chan, Chao, Charlton,
  Chassande-Mottin, Chen, Chen, Cheng, Chincarini, Chiummo, Cho, Cho, Chow,
  Christensen, Chu, Chua, Chung, Ciani, Clara, Clark, Cleva, Coccia, Cohadon,
  Colla, Collette, Cominsky, Constancio, Conte, Conti, Cook, Corbitt, Cornish,
  Corsi, Cortese, Costa, Coughlin, Coughlin, Coulon, Countryman, Couvares,
  Cowan, Coward, Cowart, Coyne, Coyne, Craig, Creighton, Creighton, Cripe,
  Crowder, Cruise, Cumming, Cunningham, Cuoco, Canton, Danilishin, D'Antonio,
  Danzmann, Darman, Da~Silva~Costa, Dattilo, Dave, Daveloza, Davier, Davies,
  Daw, Day, De, DeBra, Debreczeni, Degallaix, De~Laurentis, Del\'eglise,
  Del~Pozzo, Denker, Dent, Dereli, Dergachev, DeRosa, De~Rosa, DeSalvo,
  Dhurandhar, D\'{\i}az, Di~Fiore, Di~Giovanni, Di~Lieto, Di~Pace, Di~Palma,
  Di~Virgilio, Dojcinoski, Dolique, Donovan, Dooley, Doravari, Douglas, Downes,
  Drago, Drever, Driggers, Du, Ducrot, Dwyer, Edo, Edwards, Effler, Eggenstein,
  Ehrens, Eichholz, Eikenberry, Engels, Essick, Etzel, Evans, Evans, Everett,
  Factourovich, Fafone, Fair, Fairhurst, Fan, Fang, Farinon, Farr, Farr,
  Favata, Fays, Fehrmann, Fejer, Feldbaum, Ferrante, Ferreira, Ferrini,
  Fidecaro, Finn, Fiori, Fiorucci, Fisher, Flaminio, Fletcher, Fong, Fournier,
  Franco, Frasca, Frasconi, Frede, Frei, Freise, Frey, Frey, Fricke, Fritschel,
  Frolov, Fulda, Fyffe, Gabbard, Gair, Gammaitoni, Gaonkar, Garufi, Gatto,
  Gaur, Gehrels, Gemme, Gendre, Genin, Gennai, George, Gergely, Germain, Ghosh,
  Ghosh, Ghosh, Giaime, Giardina, Giazotto, Gill, Glaefke, Gleason, Goetz,
  Goetz, Gondan, Gonz\'alez, Castro, Gopakumar, Gordon, Gorodetsky, Gossan,
  Gosselin, Gouaty, Graef, Graff, Granata, Grant, Gras, Gray, Greco, Green,
  Greenhalgh, Groot, Grote, Grunewald, Guidi, Guo, Gupta, Gupta, Gushwa,
  Gustafson, Gustafson, Hacker, Hall, Hall, Hammond, Haney, Hanke, Hanks,
  Hanna, Hannam, Hanson, Hardwick, Harms, Harry, Harry, Hart, Hartman, Haster,
  Haughian, Healy, Heefner, Heidmann, Heintze, Heinzel, Heitmann, Hello,
  Hemming, Hendry, Heng, Hennig, Heptonstall, Heurs, Hild, Hoak, Hodge, Hofman,
  Hollitt, Holt, Holz, Hopkins, Hosken, Hough, Houston, Howell, Hu, Huang,
  Huerta, Huet, Hughey, Husa, Huttner, Huynh-Dinh, Idrisy, Indik, Ingram, Inta,
  Isa, Isac, Isi, Islas, Isogai, Iyer, Izumi, Jacobson, Jacqmin, Jang, Jani,
  Jaranowski, Jawahar, Jim\'enez-Forteza, Johnson, Johnson-McDaniel, Jones,
  Jones, Jonker, Ju, Haris, Kalaghatgi, Kalogera, Kandhasamy, Kang, Kanner,
  Karki, Kasprzack, Katsavounidis, Katzman, Kaufer, Kaur, Kawabe, Kawazoe,
  K\'ef\'elian, Kehl, Keitel, Kelley, Kells, Kennedy, Keppel, Key,
  Khalaidovski, Khalili, Khan, Khan, Khan, Khazanov, Kijbunchoo, Kim, Kim, Kim,
  Kim, Kim, Kim, King, King, Kinzel, Kissel, Kleybolte, Klimenko, Koehlenbeck,
  Kokeyama, Koley, Kondrashov, Kontos, Koranda, Korobko, Korth, Kowalska,
  Kozak, Kringel, Krishnan, Kr\'olak, Krueger, Kuehn, Kumar, Kumar, Kuo,
  Kutynia, Kwee, Lackey, Landry, Lange, Lantz, Lasky, Lazzarini, Lazzaro,
  Leaci, Leavey, Lebigot, Lee, Lee, Lee, Lee, Lenon, Leonardi, Leong, Leroy,
  Letendre, Levin, Levine, Li, Libson, Littenberg, Lockerbie, Logue, Lombardi,
  London, Lord, Lorenzini, Loriette, Lormand, Losurdo, Lough, Lousto, Lovelace,
  L\"uck, Lundgren, Luo, Lynch, Ma, MacDonald, Machenschalk, MacInnis, Macleod,
  Maga\~na Sandoval, Magee, Mageswaran, Majorana, Maksimovic, Malvezzi, Man,
  Mandel, Mandic, Mangano, Mansell, Manske, Mantovani, Marchesoni, Marion,
  M\'arka, M\'arka, Markosyan, Maros, Martelli, Martellini, Martin, Martin,
  Martynov, Marx, Mason, Masserot, Massinger, Masso-Reid, Matichard, Matone,
  Mavalvala, Mazumder, Mazzolo, McCarthy, McClelland, McCormick, McGuire,
  McIntyre, McIver, McManus, McWilliams, Meacher, Meadors, Meidam, Melatos,
  Mendell, Mendoza-Gandara, Mercer, Merilh, Merzougui, Meshkov, Messenger,
  Messick, Meyers, Mezzani, Miao, Michel, Middleton, Mikhailov, Milano, Miller,
  Millhouse, Minenkov, Ming, Mirshekari, Mishra, Mitra, Mitrofanov,
  Mitselmakher, Mittleman, Moggi, Mohan, Mohapatra, Montani, Moore, Moore,
  Moraru, Moreno, Morriss, Mossavi, Mours, Mow-Lowry, Mueller, Mueller, Muir,
  Mukherjee, Mukherjee, Mukherjee, Mukund, Mullavey, Munch, Murphy, Murray,
  Mytidis, Nardecchia, Naticchioni, Nayak, Necula, Nedkova, Nelemans, Neri,
  Neunzert, Newton, Nguyen, Nielsen, Nissanke, Nitz, Nocera, Nolting,
  Normandin, Nuttall, Oberling, Ochsner, O'Dell, Oelker, Ogin, Oh, Oh, Ohme,
  Oliver, Oppermann, Oram, O'Reilly, O'Shaughnessy, Ott, Ottaway, Ottens,
  Overmier, Owen, Pai, Pai, Palamos, Palashov, Palomba, Pal-Singh, Pan, Pan,
  Pankow, Pannarale, Pant, Paoletti, Paoli, Papa, Paris, Parker, Pascucci,
  Pasqualetti, Passaquieti, Passuello, Patricelli, Patrick, Pearlstone,
  Pedraza, Pedurand, Pekowsky, Pele, Penn, Perreca, Pfeiffer, Phelps, Piccinni,
  Pichot, Pickenpack, Piergiovanni, Pierro, Pillant, Pinard, Pinto, Pitkin,
  Poeld, Poggiani, Popolizio, Post, Powell, Prasad, Predoi, Premachandra,
  Prestegard, Price, Prijatelj, Principe, Privitera, Prix, Prodi, Prokhorov,
  Puncken, Punturo, Puppo, P\"urrer, Qi, Qin, Quetschke, Quintero,
  Quitzow-James, Raab, Rabeling, Radkins, Raffai, Raja, Rakhmanov, Ramet,
  Rapagnani, Raymond, Razzano, Re, Read, Reed, Regimbau, Rei, Reid, Reitze,
  Rew, Reyes, Ricci, Riles, Robertson, Robie, Robinet, Rocchi, Rolland,
  Rollins, Roma, Romano, Romano, Romanov, Romie, Rosi\ifmmode~\acute{n}\else
  \'{n}\fi{}ska, Rowan, R\"udiger, Ruggi, Ryan, Sachdev, Sadecki, Sadeghian,
  Salconi, Saleem, Salemi, Samajdar, Sammut, Sampson, Sanchez, Sandberg,
  Sandeen, Sanders, Sanders, Sassolas, Sathyaprakash, Saulson, Sauter, Savage,
  Sawadsky, Schale, Schilling, Schmidt, Schmidt, Schnabel, Schofield,
  Sch\"onbeck, Schreiber, Schuette, Schutz, Scott, Scott, Sellers, Sengupta,
  Sentenac, Sequino, Sergeev, Serna, Setyawati, Sevigny, Shaddock, Shaffer,
  Shah, Shahriar, Shaltev, Shao, Shapiro, Shawhan, Sheperd, Shoemaker,
  Shoemaker, Siellez, Siemens, Sigg, Silva, Simakov, Singer, Singer, Singh,
  Singh, Singhal, Sintes, Slagmolen, Smith, Smith, Smith, Smith, Son, Sorazu,
  Sorrentino, Souradeep, Srivastava, Staley, Steinke, Steinlechner,
  Steinlechner, Steinmeyer, Stephens, Stevenson, Stone, Strain, Straniero,
  Stratta, Strauss, Strigin, Sturani, Stuver, Summerscales, Sun, Sutton,
  Swinkels, Szczepa\ifmmode~\acute{n}\else \'{n}\fi{}czyk, Tacca, Talukder,
  Tanner, T\'apai, Tarabrin, Taracchini, Taylor, Theeg, Thirugnanasambandam,
  Thomas, Thomas, Thomas, Thorne, Thorne, Thrane, Tiwari, Tiwari, Tokmakov,
  Tomlinson, Tonelli, Torres, Torrie, T\"oyr\"a, Travasso, Traylor, Trifir\`o,
  Tringali, Trozzo, Tse, Turconi, Tuyenbayev, Ugolini, Unnikrishnan, Urban,
  Usman, Vahlbruch, Vajente, Valdes, Vallisneri, van Bakel, van Beuzekom,
  van~den Brand, Van Den~Broeck, Vander-Hyde, van~der Schaaf, van Heijningen,
  van Veggel, Vardaro, Vass, Vas\'uth, Vaulin, Vecchio, Vedovato, Veitch,
  Veitch, Venkateswara, Verkindt, Vetrano, Vicer\'e, Vinciguerra, Vine, Vinet,
  Vitale, Vo, Vocca, Vorvick, Voss, Vousden, Vyatchanin, Wade, Wade, Wade,
  Waldman, Walker, Wallace, Walsh, Wang, Wang, Wang, Wang, Wang, Ward, Ward,
  Warner, Was, Weaver, Wei, Weinert, Weinstein, Weiss, Welborn, Wen,
  We\ss{}els, Westphal, Wette, Whelan, Whitcomb, White, Whiting, Wiesner,
  Wilkinson, Willems, Williams, Williams, Williamson, Willis, Willke, Wimmer,
  Winkelmann, Winkler, Wipf, Wiseman, Wittel, Woan, Worden, Wright, Wu, Yablon,
  Yakushin, Yam, Yamamoto, Yancey, Yap, Yu, Yvert, Zadro\ifmmode~\dot{z}\else
  \.{z}\fi{}ny, Zangrando, Zanolin, Zendri, Zevin, Zhang, Zhang, Zhang, Zhang,
  Zhao, Zhou, Zhou, Zhu, Zucker, Zuraw, \& Zweizig}]{LIGO_first_direct_GW}
Abbott, B.~P. {et~al.} 2016, Phys. Rev. Lett., 116, 061102

\bibitem[{{Baumgarte} \& {Shapiro}(2010)}]{baumgartebook10}
{Baumgarte}, T.~W., \& {Shapiro}, S.~L. 2010, {Numerical Relativity: Solving
  Einstein's Equations on the Computer} (Cambridge University Press, Cambridge)

\bibitem[{{Baumgarte} {et~al.}(2000){Baumgarte}, {Shapiro}, \&
  {Shibata}}]{BaShSh}
{Baumgarte}, T.~W., {Shapiro}, S.~L., \& {Shibata}, M. 2000, \apjl, 528, L29,
  astro-ph/9910565

\bibitem[{{Berger}(2014)}]{Berger2014}
{Berger}, E. 2014, Ann. Rev. Astron. Astroph., 52, 43

\bibitem[{{Blandford} \& {Znajek}(1977)}]{BZeffect}
{Blandford}, R.~D., \& {Znajek}, R.~L. 1977, \mnras, 179, 433

\bibitem[{{Dietrich} {et~al.}(2015){Dietrich}, {Bernuzzi}, {Ujevic}, \&
  {Bruegmann}}]{Dietrich2015}
{Dietrich}, T., {Bernuzzi}, S., {Ujevic}, M., \& {Bruegmann}, B. 2015, ArXiv
  e-prints, 1504.01266

\bibitem[{{Dionysopoulou} {et~al.}(2015){Dionysopoulou}, {Alic}, \&
  {Rezzolla}}]{Dionysopoulou2015}
{Dionysopoulou}, K., {Alic}, D., \& {Rezzolla}, L. 2015, ArXiv e-prints,
  1502.02021

\bibitem[{{Duez} {et~al.}(2006){Duez}, {Liu}, {Shapiro}, {Shibata}, \&
  {Stephens}}]{dlsss06a}
{Duez}, M.~D., {Liu}, Y.~T., {Shapiro}, S.~L., {Shibata}, M., \& {Stephens},
  B.~C. 2006, Physical Review Letters, 96, 031101

\bibitem[{{East} {et~al.}(2016){East}, {Paschalidis}, {Pretorius}, \&
  {Shapiro}}]{PEFS2016}
{East}, W.~E., {Paschalidis}, V., {Pretorius}, F., \& {Shapiro}, S.~L. 2016,
  \prd, 93, 024011, 1511.01093

\bibitem[{{East} \& {Pretorius}(2012)}]{East2012NSNS}
{East}, W.~E., \& {Pretorius}, F. 2012, \apjl, 760, L4, 1208.5279

\bibitem[{{Eichler} {et~al.}(1989){Eichler}, {Livio}, {Piran}, \&
  {Schramm}}]{EiLiPiSc}
{Eichler}, D., {Livio}, M., {Piran}, T., \& {Schramm}, D.~N. 1989, \nat, 340,
  126

\bibitem[{Endrizzi {et~al.}(2016)Endrizzi, Ciolfi, Giacomazzo, Kastaun, \&
  Kawamura}]{Endrizzi:2016kkf}
Endrizzi, A., Ciolfi, R., Giacomazzo, B., Kastaun, W., \& Kawamura, T. 2016,
  1604.03445

\bibitem[{Etienne {et~al.}(2008)Etienne, Faber, Liu, Shapiro, Taniguchi,
  {et~al.}}]{Etienne:2007jg}
Etienne, Z.~B., Faber, J.~A., Liu, Y.~T., Shapiro, S.~L., Taniguchi, K.,
  {et~al.} 2008, Phys.Rev., D77, 084002

\bibitem[{Etienne {et~al.}(2012{\natexlab{a}})Etienne, Liu, Paschalidis, \&
  Shapiro}]{Etienne:2011ea}
Etienne, Z.~B., Liu, Y.~T., Paschalidis, V., \& Shapiro, S.~L.
  2012{\natexlab{a}}, Phys.Rev., D85, 064029

\bibitem[{Etienne {et~al.}(2010)Etienne, Liu, \& Shapiro}]{Etienne:2010ui}
Etienne, Z.~B., Liu, Y.~T., \& Shapiro, S.~L. 2010, Phys.Rev., D82, 084031

\bibitem[{Etienne {et~al.}(2012{\natexlab{b}})Etienne, Paschalidis, \&
  Shapiro}]{Etienne:2012te}
Etienne, Z.~B., Paschalidis, V., \& Shapiro, S.~L. 2012{\natexlab{b}},
  Phys.Rev., D86, 084026

\bibitem[{{Faber} \& {Rasio}(2012)}]{faber_review}
{Faber}, J.~A., \& {Rasio}, F.~A. 2012, ArXiv e-prints, 1204.3858

\bibitem[{Farris {et~al.}(2012)Farris, Gold, Paschalidis, Etienne, \&
  Shapiro}]{Farris:2012ux}
Farris, B.~D., Gold, R., Paschalidis, V., Etienne, Z.~B., \& Shapiro, S.~L.
  2012, Phys.Rev.Lett., 109, 221102, 1207.3354

\bibitem[{Giacomazzo {et~al.}(2011)Giacomazzo, Rezzolla, \& Baiotti}]{grb11}
Giacomazzo, B., Rezzolla, L., \& Baiotti, L. 2011, Phys. Rev. D, 83, 044014

\bibitem[{{Gold} {et~al.}(2012){Gold}, {Bernuzzi}, {Thierfelder},
  {Br{\"u}gmann}, \& {Pretorius}}]{gold}
{Gold}, R., {Bernuzzi}, S., {Thierfelder}, M., {Br{\"u}gmann}, B., \&
  {Pretorius}, F. 2012, \prd, 86, 121501, 1109.5128

\bibitem[{Gold {et~al.}(2014)Gold, Paschalidis, Etienne, Shapiro, \&
  Pfeiffer}]{Gold:2013zma}
Gold, R., Paschalidis, V., Etienne, Z.~B., Shapiro, S.~L., \& Pfeiffer, H.~P.
  2014, Phys.Rev., D89, 064060, 1312.0600

\bibitem[{{Gold} {et~al.}(2014){Gold}, {Paschalidis}, {Ruiz}, {Shapiro},
  {Etienne}, \& {Pfeiffer}}]{Gold2014}
{Gold}, R., {Paschalidis}, V., {Ruiz}, M., {Shapiro}, S.~L., {Etienne}, Z.~B.,
  \& {Pfeiffer}, H.~P. 2014, \prd, 90, 104030, 1410.1543

\bibitem[{Hansen \& Lyutikov(2001)}]{Hansen:2000am}
Hansen, B.~M., \& Lyutikov, M. 2001, Mon.Not.Roy.Astron.Soc., 322, 695,
  astro-ph/0003218

\bibitem[{{Hotokezaka} {et~al.}(2013){Hotokezaka}, {Kiuchi}, {Kyutoku},
  {Okawa}, {Sekiguchi}, {Shibata}, \& {Taniguchi}}]{Hotokezaka13}
{Hotokezaka}, K., {Kiuchi}, K., {Kyutoku}, K., {Okawa}, H., {Sekiguchi}, Y.-i.,
  {Shibata}, M., \& {Taniguchi}, K. 2013, \prd, 87, 024001, 1212.0905

\bibitem[{{Just} {et~al.}(2015){Just}, {Obergaulinger}, {Janka}, {Bauswein}, \&
  {Schwarz}}]{jojb15}
{Just}, O., {Obergaulinger}, M., {Janka}, H.-T., {Bauswein}, A., \& {Schwarz},
  N. 2015, ArXiv e-prints, 1510.04288

\bibitem[{Kann {et~al.}(2011)}]{Kann:2008zg}
Kann, D.~A., {et~al.} 2011, Astrophys. J., 734, 96, 0804.1959

\bibitem[{Kiuchi {et~al.}(2015)Kiuchi, Cerdá-Durán, Kyutoku, Sekiguchi, \&
  Shibata}]{Kiuchi:2015sga}
Kiuchi, K., Cerdá-Durán, P., Kyutoku, K., Sekiguchi, Y., \& Shibata, M. 2015,
  Phys. Rev., D92, 124034, 1509.09205

\bibitem[{{Kiuchi} {et~al.}(2014){Kiuchi}, {Kyutoku}, {Sekiguchi}, {Shibata},
  \& {Wada}}]{kkssw14}
{Kiuchi}, K., {Kyutoku}, K., {Sekiguchi}, Y., {Shibata}, M., \& {Wada}, T.
  2014, \prd, 90, 041502, 1407.2660

\bibitem[{{Kiuchi} {et~al.}(2015){Kiuchi}, {Sekiguchi}, {Kyutoku}, {Shibata},
  {Taniguchi}, \& {Wada}}]{kskstw15}
{Kiuchi}, K., {Sekiguchi}, Y., {Kyutoku}, K., {Shibata}, M., {Taniguchi}, K.,
  \& {Wada}, T. 2015, \prd, 92, 064034, 1506.06811

\bibitem[{{Liu} {et~al.}(2008){Liu}, {Shapiro}, {Etienne}, \&
  {Taniguchi}}]{lset08}
{Liu}, Y.~T., {Shapiro}, S.~L., {Etienne}, Z.~B., \& {Taniguchi}, K. 2008,
  \prd, 78, 024012

\bibitem[{{Lyford} {et~al.}(2003){Lyford}, {Baumgarte}, \&
  {Shapiro}}]{LBS2003ApJ}
{Lyford}, N.~D., {Baumgarte}, T.~W., \& {Shapiro}, S.~L. 2003, \apj, 583, 410,
  gr-qc/0210012

\bibitem[{{McWilliams} \& {Levin}(2011)}]{ml11}
{McWilliams}, S.~T., \& {Levin}, J. 2011, ArXiv e-prints, 1101.1969

\bibitem[{{Metzger} {et~al.}(2015){Metzger}, {Bauswein}, {Goriely}, \&
  {Kasen}}]{2015MNRAS.446.1115M}
{Metzger}, B.~D., {Bauswein}, A., {Goriely}, S., \& {Kasen}, D. 2015, \mnras,
  446, 1115, 1409.0544

\bibitem[{{Metzger} \& {Berger}(2012)}]{MetzgerBerger2012}
{Metzger}, B.~D., \& {Berger}, E. 2012, \apj, 746, 48, 1108.6056

\bibitem[{{Mochkovitch} {et~al.}(1993){Mochkovitch}, {Hernanz}, {Isern}, \&
  {Martin}}]{MoHeIsMa}
{Mochkovitch}, R., {Hernanz}, M., {Isern}, J., \& {Martin}, X. 1993, \nat, 361,
  236

\bibitem[{{Narayan} {et~al.}(1992){Narayan}, {Paczynski}, \& {Piran}}]{NaPaPi}
{Narayan}, R., {Paczynski}, B., \& {Piran}, T. 1992, \apjl, 395, L83,
  astro-ph/9204001

\bibitem[{Neilsen {et~al.}(2014)Neilsen, Liebling, Anderson, Lehner, O'Connor,
  {et~al.}}]{Neilsen:2014hha}
Neilsen, D., Liebling, S.~L., Anderson, M., Lehner, L., O'Connor, E., {et~al.}
  2014, Phys.Rev., D89, 104029, 1403.3680

\bibitem[{{Palenzuela} {et~al.}(2013){Palenzuela}, {Lehner}, {Ponce},
  {Liebling}, {Anderson}, {Neilsen}, \& {Motl}}]{PalenzuelaLehner2013}
{Palenzuela}, C., {Lehner}, L., {Ponce}, M., {Liebling}, S.~L., {Anderson}, M.,
  {Neilsen}, D., \& {Motl}, P. 2013, Physical Review Letters, 111, 061105,
  1301.7074

\bibitem[{{Palenzuela} {et~al.}(2015){Palenzuela}, {Liebling}, {Neilsen},
  {Lehner}, {Caballero}, {O'Connor}, \& {Anderson}}]{Palenzuela2015}
{Palenzuela}, C., {Liebling}, S.~L., {Neilsen}, D., {Lehner}, L., {Caballero},
  O.~L., {O'Connor}, E., \& {Anderson}, M. 2015, ArXiv e-prints, 1505.01607

\bibitem[{Palenzuela {et~al.}(2015)Palenzuela, Liebling, Neilsen, Lehner,
  Caballero, O'Connor, \& Anderson}]{Palenzuela:2015dqa}
Palenzuela, C., Liebling, S.~L., Neilsen, D., Lehner, L., Caballero, O.~L.,
  O'Connor, E., \& Anderson, M. 2015, Phys. Rev., D92, 044045, 1505.01607

\bibitem[{{Paschalidis} {et~al.}(2015{\natexlab{a}}){Paschalidis}, {East},
  {Pretorius}, \& {Shapiro}}]{PEFS2015}
{Paschalidis}, V., {East}, W.~E., {Pretorius}, F., \& {Shapiro}, S.~L.
  2015{\natexlab{a}}, \prd, 92, 121502, 1510.03432

\bibitem[{Paschalidis {et~al.}(2012)Paschalidis, Etienne, \&
  Shapiro}]{Paschalidis:2012ff}
Paschalidis, V., Etienne, Z.~B., \& Shapiro, S.~L. 2012, Phys.Rev., D86,
  064032, 1208.5487

\bibitem[{{Paschalidis} {et~al.}(2013){Paschalidis}, {Etienne}, \&
  {Shapiro}}]{Paschalidis:2013jsa}
{Paschalidis}, V., {Etienne}, Z.~B., \& {Shapiro}, S.~L. 2013, \prd, 88,
  021504, 1304.1805

\bibitem[{{Paschalidis} {et~al.}(2015{\natexlab{b}}){Paschalidis}, {Ruiz}, \&
  {Shapiro}}]{prs15}
{Paschalidis}, V., {Ruiz}, M., \& {Shapiro}, S.~L. 2015{\natexlab{b}}, \apjl,
  806, L14, 1410.7392

\bibitem[{{Ponce} {et~al.}(2014){Ponce}, {Palenzuela}, {Lehner}, \&
  {Liebling}}]{2014PhRvD..90d4007P}
{Ponce}, M., {Palenzuela}, C., {Lehner}, L., \& {Liebling}, S.~L. 2014, \prd,
  90, 044007, 1404.0692

\bibitem[{{Rezzolla} {et~al.}(2011){Rezzolla}, {Giacomazzo}, {Baiotti},
  {Granot}, {Kouveliotou}, \& {Aloy}}]{rgbgka11}
{Rezzolla}, L., {Giacomazzo}, B., {Baiotti}, L., {Granot}, J., {Kouveliotou},
  C., \& {Aloy}, M.~A. 2011, \apjl, 732, L6

\bibitem[{{Sekiguchi} {et~al.}(2015){Sekiguchi}, {Kiuchi}, {Kyutoku}, \&
  {Shibata}}]{Sekiguchi2015}
{Sekiguchi}, Y., {Kiuchi}, K., {Kyutoku}, K., \& {Shibata}, M. 2015, \prd, 91,
  064059, 1502.06660

\bibitem[{Shibata(2005)}]{ShibataPhysRevLett.94.201101}
Shibata, M. 2005, Phys. Rev. Lett., 94, 201101

\bibitem[{{Taniguchi} \& {Gourgoulhon}(2002)}]{tg02}
{Taniguchi}, K., \& {Gourgoulhon}, E. 2002, \prd, 66, 104019, gr-qc/0207098

\bibitem[{{Thornburg}(2004)}]{ahfinderdirect}
{Thornburg}, J. 2004, Class.~Quant.~Grav., 21, 743

\bibitem[{Thorne {et~al.}(1986)Thorne, Price, \& Macdonald}]{Thorne86}
Thorne, K.~S., Price, R.~H., \& Macdonald, D.~A. 1986, The Membrane Paradigm
  (New Haven: Yale University Press)

\bibitem[{{Vlahakis} \&
  {K{\"o}nigl}(2003)}]{B2_over_2RHO_yields_target_Lorentz_factor}
{Vlahakis}, N., \& {K{\"o}nigl}, A. 2003, \apj, 596, 1080

\bibitem[{{Zrake} \& {MacFadyen}(2013)}]{2013ApJ...769L..29Z}
{Zrake}, J., \& {MacFadyen}, A.~I. 2013, \apjl, 769, L29, 1303.1450

\end{thebibliography}

\end{document}